# Negative Polarization through Photon to Electron Spin Polarization Transfer in GaAs Quantum Wells

H. Kosaka[1,2], Y. Rikitake[2,3], H. Imamura[3,2], Y. Mitsumori[1,2] and K. Edamatsu[1]

[1]*Research Institute of Electrical Communication, Tohoku University, Sendai 980-8577, Japan*
[2]*CREST-JST, Saitama 332-0012, Japan*
[3]*Nanotechnology Research Institute, AIST, Tsukuba 305-8568, Japan*

We demonstrate negative polarization created by light-hole exciton excitation in g-factor engineered GaAs quantum wells measured by time-resolved Kerr rotation and polarization-resolved photoluminescence. This negative polarization is a result of polarization transfer from a photon to an electron spin mediated by a light hole. This demonstration is an important step towards achieving quantum media conversion from a photonic qubit to an electron spin qubit required for building a quantum repeater.

Quantum information can take several different forms, and it is advantageous to be able to convert between different forms. One form is photon polarization, and another is electron spin polarization.

Photons are the most convenient medium for sharing quantum information between distant locations. Quantum key distribution [1] has been demonstrated by sending photons through a conventional optical fiber up to distances around 100 km [2-5]. As the distance increases, the secure data rate decreases, due to photon loss. To expand the distance dramatically, it is necessary to realize a quantum repeater, which is based on quantum teleportation protocol [6-9]. A quantum repeater requires quantum information processing and storage, and an electron spin is a good candidate for these functionalities. Hence, we need to have quantum media conversion from a messenger photonic qubit to a processor electron spin qubit [10-12]. This requires "quantum state transfer" [13-18] from photon polarization to electron spin polarization.

The quantum state transfer is a unitary transformation between two mathematically equivalent SU(2) Hilbert spaces: from a "Poincare sphere," which describes a photon polarization state, to a "Bloch sphere," which describes an electron spin polarization state. This unitary transformation is possible only when both the "energy conservation" and the "angular momentum conservation" conditions are satisfied. The energy conservation under a magnetic field is satisfied by engineering the "electron g-factor" to zero [19-22], and the angular momentum conservation is satisfied by the spin-dependent "optical selection rule."

The optical selection rule is governed by the transfer of the angular momentum from a photon characterized by a helicity (+1 for $\sigma^+$ polarization, -1 for $\sigma^-$ polarization) to an exciton characterized by an orbital angular momentum ($L = \pm 1$). The spin-orbit coupling of a hole combines the polarization degree of freedom of a photon with the spin degree of freedom of an electron and a hole system as illustrated by the solid arrows in Fig. 1.

In this letter, we demonstrate the polarization transfer from a photon to an electron spin in a g-factor engineered quantum well, by showing "negative polarization" phenomena through the excitation of a light-hole exciton, which plays a key role in achieving the quantum state transfer as explained later [13]. The negative polarization is good proof of the angular momentum conservation in the required situation in which the electron g-factor is engineered to zero. It was demonstrated by both polarization-dependent optical pump-probe measurements and photo-luminescence (PL) measurements.

Negative polarization was first observed in a negatively charged quantum dot by resonantly excited PL measurements [23], and then by non-resonantly excited PL measurements [24]. These experiments manifested the significance of the negative polarization for optically driven spin initialization, memory, manipulation, and detection towards quantum information processing. The mechanism of their negative polarization is based on the spin flip-flop of an electron and a heavy hole mediated by the anisotropic exchange interaction [23] or spin flip of a delocalized heavy hole in a dark exciton [24]. Although the behavior is similar, the mechanism is fundamentally different in our case. The significance of our observation is that we do not need either of the two-electron composites, which suffer from the electron-electron exchange interaction [25] or strong confinement of an electron in a quantum dot.

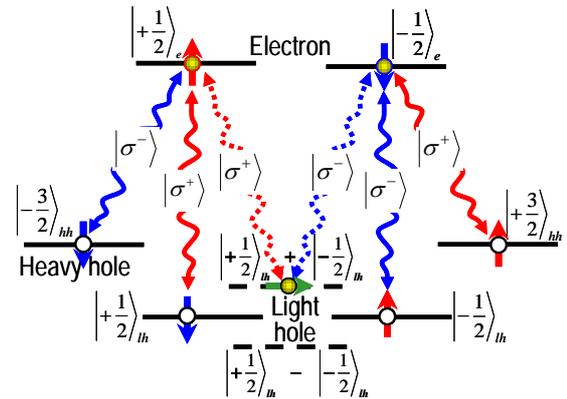

Figure 1. Schematic energy band diagram of the fabricated quantum well structure under a zero-magnetic field (solid lines) and an in-plane magnetic field (broken lines). Allowed interband optical transitions are depicted by the wavy arrows on the basis of circular polarization.

The sample we investigated contains 10 layers of 6-nm-thick GaAs quantum wells separated by 10-nm-thick $Al_{0.35}Ga_{0.65}As$ barrier layers grown by molecular beam epitaxy on a semi-insulating GaAs substrate. The sample is designed to have zero electron g-factor under an in-plane magnetic field. All measurements were performed at 5 K at normal incidence.

We first evaluated the electron in-plane g-factor by using the time-resolved Kerr rotation technique. A mode-locked Ti:Sapphire laser delivering 2-ps pulses at a repetition rate of 80 MHz was used to pump and probe the sample with a variable time delay only for the probe. After rotating the polarization of the reflected probe light with a half-wave plate, a polarization beam splitter and a Silicon balanced photodetector were used to measure the modulation of the polarization angle of the initially horizontally-polarized probe light. The polarization of the pump light was periodically altered between $\sigma^+$ and $\sigma^-$ by using a photo elastic modulator at a frequency of 42 KHz to prevent dynamic nuclear-spin polarization. Time dependence of the Kerr rotation signal revealed no indication of spin precession up to 5 T within the time range of signal decay up to around 500 ps, which implies the in-plane electron g-factor with modulus was reduced at least 20-fold from the bulk value to less than 0.02. This is the consequence of g-factor averaging over the wave function ranging in both the negatively valued GaAs well layer and the positively valued AlGaAs barrier layer. This g-factor of an electron is negligibly small compared to that of a light hole, which is estimated to be -5 for the measured quantum well, thereby assuring the assumption of an energy band diagram as shown in Fig. 1.

Next we changed the measurement setup from a one-color to a two-color time-resolved Kerr rotation system. The pulse width of the Ti:Sapphire laser was shortened to 130 fs to expand the spectral bandwidth from 1 nm to 12 nm at FWHM, and two wavelength-tunable optical band-pass filters with a bandwidth of 1.5 nm were inserted into both pump and probe paths so as to obtain pump and probe Kerr rotation spectra. The time dependence of Kerr rotation angles was measured at various pump wavelengths and at a fixed probe wavelength of 778 nm, which corresponds to the lowest heavy-hole exciton, as explained later. The initial height of a single exponential fit, excluding a coherent spike caused by the interference of pump light and probe light, is plotted in Fig. 2(a). The spectrum exhibits a remarkable negative signal at around 769 nm. The solid line in the figure is a fit to four-component Gaussian functions, and the open circles are the center wavelengths of four Gaussian components. Figure 2(b) and (c) are raw signals of time-resolved Kerr rotation measurements pumped at 769 nm and 778 nm, which are typical, respectively, for the negative and positive Kerr signals in Fig. 2(a). Figure 3(a) shows a "probe" spectrum of the Kerr rotation pumped at 769 nm, where the maximum negative signal appears in Fig. 2(a). Center wavelengths of a multi-Gaussian fit are obtained by using the same procedure as in Fig. 2(a).

The positions of peaks and dips in Fig. 2 coincide well with one of the lowest heavy-hole excitons (1HH: 778.0 nm), the second-lowest heavy-hole exciton (2HH: 771.2 nm), and the lowest light-hole exciton (LH: 767.2 nm), which are identified by the conventional non-polarized PL excitation spectrum as shown in Fig. 3(b), to a measurement accuracy around 0.5 nm. The origins of both the negative peak in Fig. 2(a) and the positive peak in Fig. 3(a) are thus assigned to the LH exciton, while the other peaks are assigned to HH excitons except for the one around 765 nm, which is around the edge of the quasicontinuum band (an electron in the lowest state and a hole in the continuum state). The pump wavelength of Fig. 3(a) is safely assigned to an LH exciton from the observations of Fig. 2(b) and Fig. 3(b).

The key signature of the above observations is the negativity or counter-polarization in a Kerr rotation signal when pumped at an LH exciton and probed at an HH exciton, which contrasts with the positive signal or co-polarization when pumped and probed at either an HH or LH exciton. The physics behind the negativity in a Kerr rotation can be understood by referring to the band configuration and the selection rules shown in Fig. 1. When an LH exciton is excited by a $\sigma^+$ photon, a spin-up electron and a spin-down hole are created simultaneously, whereas the light hole state is depolarized and relaxes to the heavy hole state much faster than the time scale of the electron spin depolarization, which is approximately 150 ps. The resulting accumulation of spin-up electrons breaks the balance of electron spin populations, thereby leading to different refractive indices for $\sigma^+$ and $\sigma^-$ light. Since the linearly-polarized probe light is a superposition of $\sigma^+$ and $\sigma^-$ light, such that $|H\rangle = |\sigma^+\rangle + |\sigma^-\rangle$ (normalization omitted for brevity), a phase delay in one component of the reflected light relative to the other results in tipping of the polarization angle. The sign, + or -, of the tip angle depends on the probe wavelength. When the probe light is tuned to an HH exciton, the $\sigma^-$ component is delayed against the $\sigma^+$ component since a spin-up electron is created only by a $\sigma^-$ photon, which is opposite to the pump polarization. The reflected light is thus phase shifted as $|\sigma^+\rangle + e^{-i\theta}|\sigma^-\rangle$, which gives a negative Kerr rotation signal when projected to the 45° linear polarization basis ($D^+$-$D^-$). In contrast, when the pump and probe lights are both tuned to either an HH or LH exciton, the reflected light is phase shifted as $e^{-i\theta}|\sigma^+\rangle + |\sigma^-\rangle$, which gives the positive Kerr rotation signal.

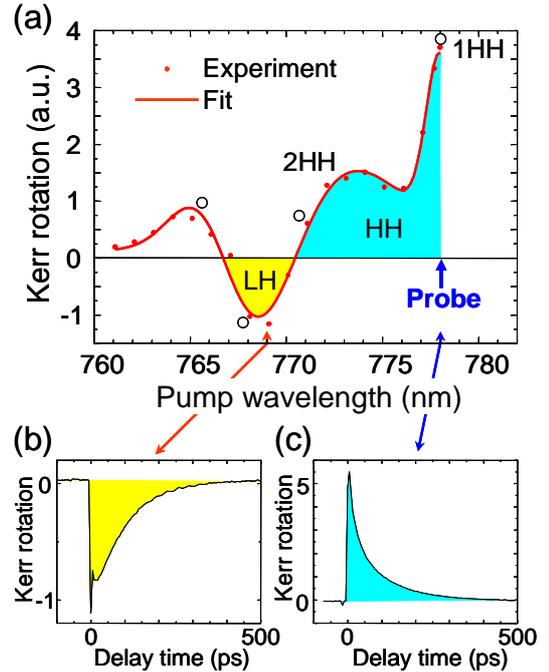

Figure 2. (a) Kerr rotation pump spectrum probed at 778 nm (1HH) at 5 K. Dots are from the initial height of an exponential fit of time-resolved Kerr rotation signals shown in (b) and (c). The solid curve in (a) is a fit to four-component Gaussian functions, whose peaks are marked by the open circles. Negative polarization appears only at the LH excitation. (b and c) Raw signal of the time-resolved Kerr rotation measurement probed at 778 nm and pumped at (b) 769 nm (LH) and (c) 778 nm (1HH).

The negativity in the Kerr rotation fundamentally originates from the "anti-parallelism" between the z-axis components of the orbital and spin angular momenta for the light hole, such that $|j=3/2, j_z=1/2\rangle$ state is decomposed into a direct product of $|l_z=1\rangle$ orbital state, and $|s_z=-1/2\rangle$ spin state. This contrasts with the positive Kerr rotation, which originates from the "parallelism" for the heavy hole, such that $|j=3/2, j_z=3/2\rangle$ state is decomposed into $|l_z=1\rangle$ and $|s_z=1/2\rangle$. As mentioned previously, the helicity of a photon is transferred only to the total orbital angular momentum of the exciton, while keeping the total spin angular momentum unchanged. The helicities of the pump and probe light to create the same electron spin state are thus opposite each other for the HH exciton and LH exciton, therefore showing negative polarization. In other words, the "negative polarization" is a manifestation of the polarization transfer from a photon to an electron spin and angular momentum conservation, especially in the case of the LH exciton.

Negative polarization is also observed in a polarization-resolved time-integrated PL spectrum (Fig. 4). PL under $\sigma^+$ excitation at normal incidence is detected through a band-pass filter tuned at 770 nm (2HH), a quarter-wave plate, and a linear polarizer. The rotation angle of the polarizer measured from a slow axis of the quarter-wave plate to achieve maximum detection intensity is plotted in Fig. 4. In ideal measurement conditions, the angle should be either +45°, which means $\sigma^+$ PL, or -45°, which means $\sigma^-$ PL. The measured values are, however, in between +45° and -45° due to background light that is linearly polarized. If we assume the background light does not depend on the wavelength being within the measured range, we can extract a degree of circular polarization from the measured angle. LH excitation leads to a negative angle, reflecting $\sigma^-$-like PL, while HH excitation leads to a positive angle, reflecting $\sigma^+$-like PL. The origin of this negativity is thought to be the same as that in the Kerr rotation. In this measurement, we tuned the detection wavelength to the 2HH exciton because its decay time is presumed to be much shorter than that of a 1HH exciton, so as to efficiently extract the electron spin information before losing its coherence.

In both the Kerr rotation and the PL spectra, visibility of the LH signal does not seem as high as that of the HH signal. This is partially because of the wide bandwidth of the filter used in the current measurements. Another reason might be an energy overlap of the HH band or the quasicontinuum band on the LH band because of a non-zero k component or inhomogeneity arising from the multi-quantum wells. This overlap can be lifted to obtain higher visibility by confining the exciton to a single quantum dot. Another way to eliminate the effects of HH is to introduce a tensile strain to make LH lower in energy than HH.

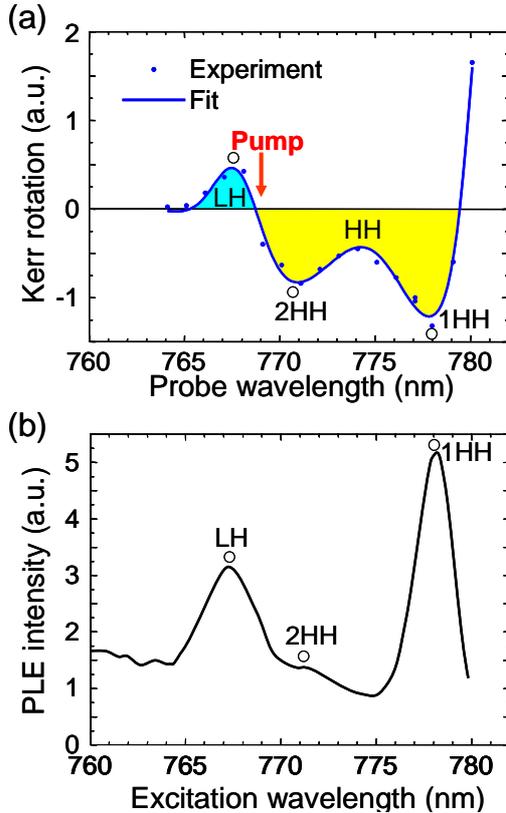

Figure 3. (a) Kerr rotation probe spectrum pumped at 769 nm (LH) at 5 K. Dots, a solid curve and open circles were obtained using the same procedure as in Fig. 2(a). (b) Conventional non-polarized PL excitation spectrum of the same wafer as a reference.

For the next important step, we demonstrate the "coherence transfer" of polarization from a photon to an electron spin. Coherence transfer is an extension of "polarization transfer," where only the z-axis projection component in a Poincare-Bloch sphere is transferred. Here we need to apply an in-plane magnetic field to the system in order to lift the spin degeneracy of the light hole to reconfigure bonding and anti-bonding states, which enables selective excitation of a light hole in one spin eigen state (bonding state). This bonding state makes it possible to configure the V-type three-level system as shown in Fig. 1 by the broken arrows, thus transferring the arbitrary superposition state of a photon to that of an electron spin via photo absorption. This coherence transfer guarantees the quantum state transfer from a photon to an electron spin. Another step toward developing a quantum repeater is to achieve a long spin memory time, which is possible by separating an electron from a hole before a hole spin degrades the electron spin coherence via the exchange interaction [26].

In summary, we have observed negative polarization in both time-resolved Kerr rotation and polarization-resolved PL spectra measurements when an LH exciton is pumped (excited) and an HH exciton is probed (detected). This negative polarization originates from the polarization transfer from a photon to an electron spin mediated by a light hole, which provides a V-type transition that is essential for enabling the coherent state transfer. This demonstration will be an important step toward realizing quantum media conversion from a photonic qubit to an electron spin qubit, which is required for building a quantum repeater or an interface for a distributed parallel quantum computer. We have examined a GaAs-based quantum well structure, which is favorable for electrical quantum transport devices to manipulate a single electron spin [12] or two electron spins [11] in gate-defined quantum dots. Although the wavelength of photo absorption in GaAs is about half of the optical communication wavelength, current optical up-conversion technology is efficient enough to convert the wavelength of a photonic qubit from 1.5 µm to around 770 nm.

We are grateful to T. Takagahara for his valuable discussions. This work was supported by CREST-JST, ERATO-JST, and SCOPE.

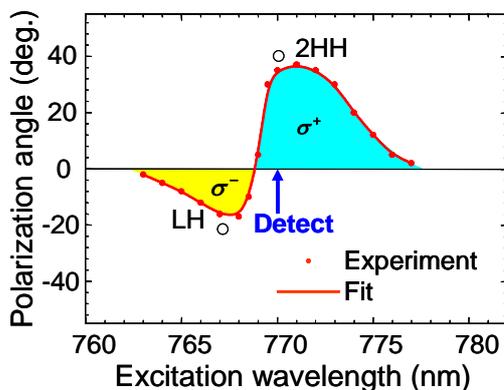

Figure 4. Polarization-resolved PL spectrum detected at 770 nm (2HH). Dots were obtained by searching the angle of polarizer rotation to obtain maximum PL intensity detected through a quarter-wave plate and a linear polarizer under a $\sigma^+$ pump.